\documentclass[aps,twocolumn,twoside,floatfix,nofootinbib,a4paper]{revtex4}
\usepackage{graphicx}
\usepackage{marginnote}

\usepackage{hyperref}
\usepackage{enumitem,kantlipsum}
\usepackage[usenames,dvipsnames]{color}
\usepackage{amsmath}
\usepackage{amssymb}
\usepackage{amsfonts}
\usepackage{mathrsfs}
\usepackage{bm}
\usepackage[all]{xy}
\usepackage{hyperref}
\usepackage{tikz-cd}

\newcommand{\beq}{\begin{equation}}
\newcommand{\eeq}{\end{equation}}
\newcommand{\bea}{\begin{eqnarray}}
\newcommand{\eea}{\end{eqnarray}}
	
\newcommand{\ea}{\end{align*}}

\newcommand{\bma}{\begin{pmatrix}}
\newcommand{\ema}{\end{pmatrix}}

\usepackage[percent]{overpic}
\usepackage{overpic}

\begin{document}
\title{Little Red reionization factories}
\author{Fan Zhang} 
\affiliation{Institute for Frontiers in Astronomy and Astrophysics, Beijing Normal University, Beijing 102206, China}
\affiliation{Gravitational Wave and Cosmology Laboratory, School of Physics and Astronomy, Beijing Normal University, Beijing 100875, China}
\affiliation{Advanced Institute of Natural Sciences, Beijing Normal University at Zhuhai, Zhuhai 519087, China}
\date{\today}

\begin{abstract}
\begin{center}
\begin{minipage}[c]{0.9\textwidth}
In this note, we suggest the possibility that an elucidation of the nature of the numerous Little Red Dots (LRDs) at high redshifts, may facilitate the resolution of another concurrent cosmic puzzle, namely reionization. Specifically, it is hypothesized that intergalactic HI gas is compelled by the tidal field associated with a growing gravitational entropy, in the form of gas streams, into colliding at the LRD sites. The resulting shock heating and compression encourage starburst activities, which subsequently photo-ionize the gas into HII, that in turn shines into the rest-optical via bremsstrahlung radiation within the minimal-timescale-cutoff (i.e., red) regime, before escaping back into the intergalactic space, propelled by the newly fusion-injected energy. 
\end{minipage} 
 \end{center}
  \end{abstract}
\maketitle

%\raggedbottom
\section{Introduction \label{sec:Intro}}
The red rest-optical spectra of Little Red Dots (LRDs) \cite{2023Natur.616..266L,2024ApJ...963..129M} often terminate at strong Balmer breaks (see e.g., \cite{2024ApJ...969L..13W,2025ApJ...984..121W,2024ApJ...964...39G,2024Natur.628...57F,2024ApJ...977L..13B,2024ApJ...975..178K}), and a large (possibly still resolution-limited) percentage of these objects exhibit Balmer absorption lines \cite{2024ApJ...963..129M,2024A&A...691A.145M,2024MNRAS.535..853J,2025ApJ...986..126K,2024ApJ...974..147L,2025arXiv250311752D}. Because the $n=2$ state of HI is short-lived, the robust detection of such features thus imply the ubiquitous presence of rather isotropic extremely dense HI gas, with number density $n_{\rm H} \gtrsim 10^9 \, \text{cm}^{-3}$ \cite{2024MNRAS.535..853J}.  
Possible concentrates of HI around radiating energy sources, during the relevant cosmic epoch\footnote{Indirect observation-constrained ionization fraction rises rapidly between redshifts $z\sim 6$ and $8$, see Fig.~5 in \cite{annurev:/content/journals/10.1146/annurev-astro-120221-044656} and references therein. This period overlaps well with the interval when fitted LRD number density peaks, see Fig.~1 in \cite{2025ApJ...988L..22I}.}, in objects so populous that they are easily found within the James Webb Space Telescope's narrow field of view, would necessarily have to be scrutinized more carefully in the context of reionization. Moreover, observational evidences suggest that reionization occurred in a rather abrupt phase-transition-like fashion (see e.g., Fig.~10 in \cite{2022LRCA....8....3G} and references therein), that may be difficult to model by simply extrapolating lower redshift astrophysics, and LRDs represent a convenient quasi-transient early-Universe feature that could be enlisted for this task. 

We propose a speculative scenario where the LRDs play a role in reionization (the hypothetical objects we propose will be abbreviated as $\iota$LRDs below, to distinguish them from the observational concept of LRDs), likely in conjunction with more traditional channels, making the two phenomena not only concurrent but also concomitant. Specifically, we envision that:
\begin{enumerate}
\item[$\mathfrak{R}1.$]
The primordial (He and other elements admixed) HI gas is driven into intergalactic scale streams\footnote{Cf., ``dark flows'' \cite{2008ApJ...686L..49K} on larger scales.} by the gravitoelectric/tidal field of the intrinsically gravitational degrees of freedom (carried within the Weyl curvature \cite{galaxies7010027}, hereafter referred to as the Cosmic Tide), that would escape scrutiny in $N$-body simulations accounting for only mutual Coulomb attractions between matter particles. The Weyl freedom represents an important component of cosmic entropy \cite{Penrose:1980ge}, and grows along the thermodynamic arrow of time only to a sufficient strength to drive the colliding streams into sufficient relative speeds ($\sim 40\, \text{km}\,\text{s}^{-1}$, see Sec.~\ref{sec:photoio} below) at $z\sim 8$, for the purpose of promoting star formation.

\item[$\mathfrak{R}2.$]
Specifically, when sufficiently fast such streams smash together, the shock heating of atoms pushes temperature well above the Lyman-$\alpha$ (Ly$\alpha$) radiation threshold. So atomic radiative cooling is available to continue the effort started by inverse-Compton scattering against the Cosmic Microwave Background (requiring $z>6$, see Sec.~\ref{sec:photoio} below), and depress temperature more quickly than collisionally ionized hydrogen can recombine, leaving an abundance of the charged species to catalyze the formation of molecules that can effect further cooling, leading eventually to population III star formation. These massive and hot stars emit copious amounts of Lyman continuum (LyC), as well as other hard ionizing (of heavier elements) photons. 

\item[$\mathfrak{R}3.$]
The photo-ionized (much of the shock-collisionally ionized hydrogen recombined during cooling and is now ionized once again) HII then escapes back into the intergalactic space along the interstitial directions of the incoming streams, propelled by stellar energy injection, as well as the Cosmic Tidal field. Namely, while the accumulation of baryonic matter builds up a gravitational potential well, photo-heating and tides (together with supernovae whenever present) nevertheless helps hot HII gas climb out.

\end{enumerate}
In summary, the reionization being contrived here does not happen in situ (i.e., there is no need to accomplish the perilous escape of photons into the intergalactic space), but occurs within compact pass-through rapid-processing facilities. 

On the $\iota$LRD phenomenology, we note that (citations are for observational correlatives, that are assigned different explanations in those works; we only reference the raw phenomenological features and make no attempt at being exhaustive with the vast amount of fast evolving literature):
\begin{enumerate}
\item[$\mathfrak{L}1.$]
The charged species within HII emit bremsstrahlung radiation much like in the HII regions surrounding later-type stars, which has a flat-into-rapid decline (as frequency increases) spectrum for each pulse encounter, because there is a minimum timescale in the scattering problem, determined by how fast the electron could traverse the impact parameter, which is the minimal distance between the charges. After integrating over admissible impact parameters and the kinetic energy distribution of the electrons, acquired through photo-heating and the subsequent thermalization, one obtains a red spectrum in the rest-optical observation window. This reddening can be quite steep as it essentially implements a cutoff. In the regular stellar HII regions, this sector of the spectrum is beyond radio frequencies, thus is seldomly shown and may not be what immediately comes to mind when one thinks of HII region bremsstrahlung spectra. In particular, the reddening in the single-scatter profile is unimportant for radio, thus easily forgotten. 

The spectrum transitions into a rollover (cf., e.g., \cite{2024ApJ...968....4P,2025arXiv250302059S}) at rest-infrared, attributed to the opacity of the HII region becoming significant. Further into longer wavelength, the spectrum slopes back down according to the Planck's law profile. This segment typically also appears in the radio spectrum of late-type stellar HII regions, but is shifted into higher (infrared) frequencies within our context because of the higher emission measure associated with a galactic-sized much denser HII region. 
In summary, within the present proposal, the gas, and not (directly) the stars or active galactic nuclei (AGN) accretion disks, is the radiating agent for the longer wavelengths. 

The $\iota$LRD ultraviolet continuum is stellar in origin. However, the high energy radiation from the central population III stars needs to penetrate thick and dense layers of various matter screens such as HI and possibly some dust; while the continuum is also typically observed at the lower energy side, biased in favor of lighter cooler sources. Therefore, lighter stars in the outer (cf., e.g., \cite{2024A&A...691A..52K,2025arXiv250520393Z,2025arXiv250509542T,2025ApJ...983...60C}), lower gas density, regions are likely more important for this segment of the spectrum. These stars may be of population II, seeded by metals assembled in earlier short-lived population III stars when they were alive, and kicked outwards by supernovae when they died.  

\item[$\mathfrak{L}2.$]
When there are significant recent supernova activities (does not always happen, especially as some stars collapse directly into black holes, cf., e.g., \cite{2024arXiv241114383R} for observational cases of only narrow lines; but population III stars are short-lived, so if the mass range is appropriate for supernovae, they would explode promptly and their signatures can appear concurrently with other $\iota$LRD phenomenologies), the ejecta would ram into ambient HI gas and transfer their kinetic energy onto the atoms, taking them up onto excited energy levels. The excited HI is then eligible for emitting Balmer H$\alpha$ and H$\beta$ lines (cf., e.g., \cite{2024ApJ...964...39G}). The ejecta also bequeath momentum onto neutral hydrogen, thereby broaden the Balmer emission lines into large and diverse (there is a scatter of stellar masses and the ejecta slows down as it propagates) widths reaching into thousands of $\text{km}\,\text{s}^{-1}$ (cf., e.g., \cite{1997ARA&A..35..309F} and \cite{Greene_2024}; note supernovae by population III stars could be even more energetic than for late-type stars). The supernovae ejecta are also rich in star-baked metals that could produce, e.g., broadened $[{\rm O_{III}}]$ lines (cf., e.g., \cite{2024A&A...691A..52K}). 

Elsewhere, the photons emitted by stars that are below LyC in energy but sits elsewhere on the Lyman series (which is discrete, thus this photon population is limited), are also of high frequency so can transit the intervening HII plasma unscathed, reaching then into the HI regions closest to the stars to excite hydrogen atoms. The subsequent Balmer emission lines would be narrow. In other words, there can be narrow and broad superposed dual components in the lines, cf., e.g., \cite{2024A&A...691A..52K}. 

\item[$\mathfrak{L}3.$] 
The high density cooler HI gas in the outer periphery contains a population of collisionally excited atoms sitting on the $n=2$ state, (almost or fully) saturated at the equilibrium fraction. 
While participating in bulk motion along our line-of-sight, within the incoming streams or for better observational prospect, their more isotropic post-collision forward-sprays and/or splash-backs, they could account for the observed slightly red and/or blue-shifted Balmer absorption lines, cf., e.g., \cite{2025arXiv250710659L,2025ApJ...980L..27I,2024ApJ...974..147L,2025arXiv250113082J,2025arXiv250614870D}. The fact these absorption lines can be narrow yet shifted, is consistent with the hydrogen atoms engaging mostly in bulk and not thermal motion. 

The high energy tail of the bremsstrahlung radiation continuum is also intercepted and redirected into ionizing the $n=2$ neutral hydrogen. This process double-taps the ultimately stellar energy injection for ionization, and results in prominent Balmer breaks. The breaks often mark the bottoms of the V-shaped spectra, where the optical bremsstrahlung emission gives way to the stellar ultraviolet contribution, unless a strong stellar flux has already overpowered bremsstrahlung at longer wavelengths.  

\item[$\mathfrak{L}4.$] 
The distribution of matter coming through a violent collision process would typically be rather misshapen (also temporally variable but on long dynamical timescales of parsecs over tens to hundreds of kilometers per second, i.e., $\gtrsim 10^4\, \text{yr}$, thus difficult to discern, cf., e.g., \cite{2024arXiv240704777K,2025arXiv250710659L}), in contrast to more gradual accretion processes where relaxation timescales are shorter than those of galaxy growth, allowing for leveling and smoothing. In other words, $\iota$LRDs are expected to be irregular in shape. The photometric analysis of resolved LRDs indeed reveal highly disturbed spatial morphology, especially off-centered asymmetric blobs, that could plausibly be direct visuals on the post-collision forward-sprays and/or splash-backs, cf., e.g., \cite{2024arXiv241114383R,2025ApJ...983...60C,2024A&A...691A..52K}.  

\end{enumerate}
The proposal above invokes no black holes and accretion flows around them to account for the prominent features of the spectrum, thus reconciles well with the lack of strong X-rays emissions (see e.g., \cite{2024ApJ...974L..26Y,2024ApJ...963..129M,2025arXiv250316600D}), mid-IR bumps (see e.g., \cite{2024ApJ...968...34W,2024ApJ...968....4P}) or radio signals (see e.g., \cite{2024arXiv240610341A}) seen from the LRDs. Also, because the rest-optical spectrum is intrinsically very red, there is no need for dust-reddening as with the traditional evolved stars alternative, see e.g., \cite{2025arXiv250302059S,2025arXiv250301945X,2025arXiv250300998A,2025arXiv250316600D} for observational evidence against the substantial presence of dust.

In the next few sections, we examine in greater detail, several of the more requisite and distinctive fundamental ingredients. Because of the complexity involved in each and every step of the chain of physical processes, we can but provide merely a stickman caricature of the $\iota$LRDs, with all estimates based off representative rough mid-values (fleshing out the extremities of parameter ranges would likely require examining atypical processes), that are subject to considerable uncertainty and doubtlessly future revisions. 

\section{Cosmic Tide \label{sec:tide}}
Einstein's equations and Bianchi identities together form a system of equations governing the evolution of spacetime geometry, in particular the gravitational dynamics that drive structure formation, and this system can be recasted into a form resembling Maxwell's equations (see e.g., \cite{2011PhRvL.106o1101O}). Analogous to the situation with electromagnetism, matter provides the source terms, but there are also intrinsic dynamics within the field itself (including, but not limited to, gravitational waves, analogous to electromagnetic waves). Structure formation simulations accounting for only Coulomb attraction between matter particles, is akin to simulating a dynamical plasma evolution but only acknowledging mutual Coulomb electric interactions between the particles, with the transformations within the electromagnetic field itself (e.g., displacement currents etc) disregarded. There could thus be some features being missed (e.g., those analogous to plasma waves, that blends particle motions with the intrinsic field evolvements). 
 
Within General Relativity, the Riemann tensor breaks down into a Ricci half and a Weyl half\footnote{The former is the average of sectional curvatures (Gauss curvature of geodetic surfaces developing out of a base point), while the latter represents their differences \cite{1987GReGr..19..771H}, and thus require anisotropy to thrive.}, with the former equating to matter stress-energy (after trace reversal), but the latter can be present in vacuum (and indeed even when there are no matter sources anywhere at all), and carries within it\footnote{In the residual resulting from subtracting off an artificial Coulomb component, directly tied to the instantaneous distribution of matter.} the intrinsic gravitational field degrees of freedom. It was hypothesized \cite{Penrose:1980ge,2010JPhCS.229a2013T,2018FoPh...48.1177P} that this freedom represents an additional, and in fact dominant, source of entropy in the Universe, and its absence in a homogeneous and isotropic geometry is the reason why the big bang represents a low entropy start to the Universe, even though the matter content is already in an entropy-maximizing thermal equilibrium. As the Universe evolves then, the awakening of this intrinsic gravitational freedom (seeded by anisotropies reflected in the Cosmic Microwave Background), and its associated entropy growth, gives headroom for the subdominant entropy in the matter sector to decrease, and thus for matter to clump up (higher entropy matter tends to spread out more uniformly, e.g., ideal gas in a container). In other words, structure formation is fundamentally made possible by the switching-on of the intrinsic field freedom, carried within the Weyl tensor. Therefore, the omission of this intrinsic field dynamics is not insignificant. 

At present, a fully relativistic structure formation simulation is not yet available, especially as the spatial boundary conditions are not certain (e.g., the spatial topology of the Universe is not fully determined). Nevertheless, we could qualitatively assess what new features might arise when we add appreciable intrinsic freedom contributions by first ignoring the matter terms (e.g., assuming matter is tenuous, as appropriate for intergalactic spaces). The spacetime curvature approach to gravitoelectromagnetism (see e.g., \cite{2003gr.qc....11030M}), shows that over a Fermi coordinate system $(T,{\bf X})$ (the qualitative conclusions below are robust against gauge changes though) surrounding the worldline of a reference observer, as the spatial origin, we can write the relative acceleration for a nearby particle in a Lorentz form
\bea
\frac{d^2 {\bf X}}{dT^2} = -{\bf E} -2 {\bf V}\times {\bf B}\,. 
\eea
For now we concentrate on the gravitoelectric influence (because the gravitomagnetic ${\bf B}$ contribution involves speeds, both in its sourcing and its action on test particles, that are normalized by the speed of light and thus small) prescribed by 
\bea \label{eq:EDef}
E_i(T,{\bf X}) = R_{0i0j}(T) X^j + \cdots\,,
\eea
where the twice temporal projected Riemann tensor $R_{0i0j}$ is the familiar tidal/tendex \cite{2011PhRvL.106o1101O} tensor, which at each location, including at the reference origin point, is a $3\times 3$ traceless matrix. The traceless feature dictates that it must contain eigenvalues of different signs. We can of course choose the Fermi spatial coordinates to align with the eigenvectors of this tidal tensor, in which case it is obvious that particles displaced from the origin along the eigenvector directions corresponding to positive eigenvalues will be pulled towards the origin, while those along negative eigenvalue directions will be repelled (those in the intervening directions move according to an eigenvalue-weighted proximity to the eigenvectors). 

Let the reference point be the center of an $\iota$LRD, then we would see nearby primordial gas being accelerated towards it along some directions. Specifically, there can be a maximum of two eigenvectors with positive eigenvalues, and corresponding to each eigenvector, there are two headon-colliding streams approaching from either side of the $\iota$LRD center. Furthermore, if needed, the repelling direction(s) can also help extract gases out of $\iota$LRDs. The $\iota$LRDs thus effectively function as saddle nodes in a cosmic gas circulation network driven in part by the intrinsic gravitational field degrees of freedom. Astrophysical processes also help establish the pattern and character of the flows. In particular, because of the violent fusion physics occurring at these nodes, to be discussed in the next section, that inject energy, the gas flowing through becomes ionized, turning from the primordial neutral gas into the ionized intergalactic medium that we observe today. 

In order to gauge the strength of the Cosmic Tidal acceleration $\bf{E}$, we recall that there exists an empirical critical acceleration scale of $a_0 \approx 10^{-10}\, \text{m}\, \text{s}^{-2}$, that despite being first identified by the Modified Newtonian Dynamics theory \cite{1983ApJ...270..365M}, can also be assigned a purely phenomenological role independent of the physics of that theory, as a marker flagging at what strength the Newtonian Coulomb attraction from baryonic matter needs to be padded/enhanced in order to account for the rotation curves in the outer regions of galaxies, as well as the interactions between nearby galaxies. The Cosmic Tidal field provides an alternative physical interpretation for $a_0$. Specifically, $\bf{E}$ exists as an omnipresent background, that is overwhelmed in the interiors of galaxies by the matter-sourced Coulomb field, but surfaces in the outer regions and intergalactic space as soon as the Coulomb competitor declines into a more feeble strength than itself. With this assumption, one can predict that, over the intergalactic distance scale of $\sim \text{Mpc}$, neutral hydrogen atoms would be tidally accelerated by $a_0 \sim \overline{|\bf{E}|}$ into a maximum speed on the order of $\sim 10^3 \,\text{km}\, \text{s}^{-1}$, when they hit the saddle point sites where $\iota$LRDs reside. Note though, the spacing between structures and $a_0$ both evolve with redshift in ways that we cannot yet quantitatively constrain, either theoretically or observationally, so the estimate above is merely a rough indication.

\section{Photo-ionization \label{sec:photoio}}
There is not expected to be any prior baryonic concentration and interactions at the saddle nodes before the streams arrive, so metals would be lacking and the only ingredient for building the stars are the primordial elements, thus the stars emerging must be of population III. Although massive population III stars do not emit many more LyC photons throughout their lives than later-types \cite{2023ARA&A..61...65K}, because of the overall shorter lifespan, they nevertheless do so in much shorter time intervals, on the order of a few million years for a $20 \text{M}_{\odot}$ to $120 \text{M}_{\odot}$ mass range \cite{2023ARA&A..61...65K}. This contrasts with the billions of years that a solar-mass late-type star would shine for. For the $\sim 300$ million years between redshifts $6$ and $8$, there could have been many generations of population III stars (until metallicity builds up sufficiently so population II formation takes over), while a late-type star would have only dumped a fraction of its total LyC photons. Consequently, if one wishes to achieve a rapid phase-transition style reionization, then population III stars are preferable, despite being more difficult to make. 

Fortunately, at saddle node locales where the gas streams collide, conditions may be suitable for population III star formation, as the tidally accelerated supersonic streams of gas would smash together there, leading to shockwaves with the associated compression and heating (enabling subsequent radiative cooling) in the postshock collision-disrupted flow. We turn now to more details. 

\subsection{Massive stars}
The stellar formation efficiency is decided by the winner in the local tug-of-war between gravitational collapse and pressure (or rotational) support. At the saddle node sites, where HI gas have been channeled in to build up into higher density, baryonic matter dominates, we can therefore lean on traditional gravitational collapse theories, without needing to account for the intrinsic field freedoms. 
We begin by noting that the shock from colliding streams realizes a more extreme version of the cloud-cloud collision scenario during infall into more traditional gravitational potential wells, that allows for a steady state modeling of the shock \cite{1987ApJ...318...32S}. 
We can therefore import standard expressions \cite{1978ppim.book.....S,1987ApJ...318...32S} (quite conveniently for us, expressed in the frame of the shock, which is the $\iota$LRD rest frame) 
\bea
T_{\rm post} = \frac{\mu_{\rm pre}}{\rho_{\rm post} }\frac{ p_{\rm post}}{k_B}\,,
\eea
where $\mu_{\rm pre}$ is the preshock average mass per particle, and the mass density is given by ($v_{\rm sh}$ is the shock speed, i.e., the relative speed between shock front and upstream/preshock matter, and $\gamma$ is the adiabatic index)
\bea
\frac{\rho_{\rm pre}}{\rho_{\rm post}} = \frac{v_{\rm post}}{v_{\rm sh}} = \frac{(\gamma -1)+\frac{2}{M^2_{\rm s}}}{\gamma+1}\,,
\eea
with 
\bea
M_{\rm sh} \equiv \left(\frac{\rho_{\rm pre} v^2_{\rm sh}}{\gamma p_{\rm pre}}\right)^{1/2}
\eea 
being the Mach number of the shock. We see that indeed, a supersonic shock would increase gas density and temperature (the collision essentially randomizes the bulk motion into thermal motion). In particular, in the large Mach number limit, we have $n_{\rm post} \sim 4 n_{\rm pre}$ and $T_{\rm post} \sim 3 \mu_{\rm pre} v^2_{\rm sh}/(16 k_{\rm B})$ \cite{2002ApJ...569L..61Q,1987ApJ...318...32S}. 
On the other hand, the pressure ratio is given by 
\bea
\frac{p_{\rm post}}{p_{\rm pre}} = \frac{2\gamma M^2_{\rm s}-(\gamma -1)}{\gamma+1}\,,
\eea
so we see that supersonic shocks do not necessarily reduce pressure directly. Nevertheless, the temperature increase opens us a radiative cooling pathway that eventually leads to a loss of pressure so gravity could more easily overpower, and stars readily form.

This cooling is a multi-staged process. One first notes that the shock may achieve nearly full collisional ionization of hydrogen at high enough $T_{\rm post}$, specifically when $k_{\rm B} T_{\rm post} > h \nu_{\rm ion} \equiv 13.6\, \text{eV}$, giving $T_{\rm post} \gtrsim 10^5\, \text{K}$ or equivalently $v_{\rm sh}$ just below $100\,\text{km}\,\text{s}^{-1}$ in the $M_{\rm sh} \gg 1$ case. The ionized hydrogen would not immediately recombine when above a few multiples of $\sim 10^4 \, \text{K}$ \cite{1987ApJ...318...32S,1987RMxAA..14...58S}, because the (Case-A) recombination rate per proton is given by $n_e \alpha_{\rm HII}$ \cite{2022LRCA....8....3G}, where $n_e$ is the proper electron density and 
\bea \label{eq:RecomRate}
\alpha_{\rm HII} \propto \left(\frac{T_{\rm post}}{10^4 K}\right)^{-0.7} \, \text{cm}^3 \,\text{s}^{-1} \,,
\eea
leveling out at low values at above $10^4 \, \text{K}$. 
In fact, when $T_{\rm post}$ lands well above the threshold $T_{\alpha} \sim 7000-8000\,\text{K}$ for Ly$\alpha$ emission, a non-equilibrium situation where atomic cooling acting faster than recombination is achieved, so large free electron and proton populations survive at $T_{\alpha}$ \cite{1986ApJ...302..585M,1987ApJ...318...32S}, beyond what the equilibrium abundance at this temperature would otherwise predict. These charged particles in the denser regions in turn catalyze the formation of $\text{H}_2$ via the gas-phase $H^-$ and $H^+_2$ reactions (also helped by Ly$\alpha$ cooling quasi-isothermally driving up gas density \cite{2002ApJ...569..558O}), which subsequently radiate in its rotational-vibrational lines and further cool the gas isobarically towards $\sim 10^2\, \text{K}$, via a positive feedback where cooler temperature encourages additional $\text{H}_2$  formation and vice versa. The cooling drives up density along the way, towards gravitational instability, as density is inversely proportional to temperature during isobaric compression, while Jeans mass scales inverse-quadratically to density. This sequence of events realizes a non-standard Pop III.2 pathway for the formation of population III stars \cite{2023ARA&A..61...65K}, that have already been carefully scrutinized by e.g., \cite{1986ApJ...302..585M,1987ApJ...318...32S,1987RMxAA..14...58S} for other shocking mechanisms.   

Fortuitously as well, the photo-dissociation effect of Lyman-Werner background from other stars becomes less disruptive to the $\text{H}_2$ molecular cooling mechanism when initial cooling is atomic \cite{2014MNRAS.445..544S,2015MNRAS.451.2082G,2015MNRAS.453.2901G,2010MNRAS.402.1249S,2015MNRAS.446.3163L,2015MNRAS.452.1233H}, due e.g., to sufficient density achieved that self-shield against these photons. Therefore, this pathway is not shut down after a starburst gets underway, and we do not witness an early quenching of pop III star formation (cf., e.g., \cite{2002ApJ...569L..61Q}; also note that higher $v_{\rm sh}$ leads to higher final $\text{H}_2$ concentration in the presence of background radiation \cite{1987ApJ...318...32S}).  

\subsection{Epochal attendance}
There is a particularly interesting observation made in \cite{1986ApJ...302..585M}, that at $v_{\rm sh} \gtrsim 200 \, \text{km}\, \text{s}^{-1}$, a radiative shock fails to form (fully ionized post-shock matter taking too long to cool so atomic hydrogen remains absent) unless $z>6$, when the inverse-Compton scattering against the Cosmic Microwave Background is able to step in (its characteristic cooling time is $\propto (1+z)^{-4}$). This additional energy leakage from the gas initiates the cooling sequence (as verified by \cite{1987ApJ...318...32S}), until temperature drops sufficiently (see Eg.~\ref{eq:RecomRate}) that HI line emission takes over. If the tidal streams indeed typically collide with a speed exceeding this threshold of $200 \, \text{km}\, \text{s}^{-1}$ at lower $z$ (recall $\overline{|{\bf E}|}$ grows gradually over time), then we have an explanation why the LRD population exhibit a downturn at $z\sim 6$. Simultaneously, the rate of ionization would dive if $\iota$LRDs contribute dominantly, marking an apparent end to the (rapid) reionization era, with a generically nonvanishing residual neutral fraction. Alternatively, the hydrogen neutral fraction might have already hit, and become stuck at, the (almost) zero floor at a higher redshift (the observational constraint for neutral fraction between $z\sim 6-7$ are only upper limits, see Fig.~5 of \cite{annurev:/content/journals/10.1146/annurev-astro-120221-044656}), which would be the case if ionization by $\iota$LRDs (likely also helped by other mechanisms) is so highly efficient that they get the job done without even needing to enlist the trailing regiment of lower redshift $\iota$LRDs. 

On the other end, it is additionally noted that star formation also shuts down when $v_{\rm sh} \lesssim 20 \, \text{km}\, \text{s}^{-1}$ \cite{1986ApJ...302..585M,1987RMxAA..14...58S}, as the shock becomes too weak to sufficiently ionize hydrogen to drive molecule formation. As the Cosmic Tidal force (gravitational entropy) builds up only gradually over time, the streams they drive would therefore only pick up speed at sufficiently low $z$. Therefore, one would expect a sharp turn-on of $\iota$LRDs, and simultaneously their reionization effort, presumably at $z\sim 8$ (relativistic cosmological simulation may be required for a quantitative match). At other times, even close to the present, rarer cases when significant cold intergalactic matter deposits exist to be collected into streams, and the sweet-spot interval $20 \, \text{km}\, \text{s}^{-1} \lesssim v_{\rm sh}  \lesssim 200 \, \text{km}\, \text{s}^{-1}$ (with appropriate adjustments according to the preshock ionization fraction changes and metal contamination in the intergalactic medium, as relevant for post-reionization epochs) is chanced upon, may present us with additional sprinkles of $\iota$LRDs. It may be of interest (but not statistical conviction yet) to note that the blue and redshifts of the Balmer absorptions lines for low $z$ LRDs found by e.g., \cite{2025arXiv250710659L} indeed fall within this opportune $v_{\rm sh}$ range, while higher $z$ measurements obtained by e.g., \cite{2025ApJ...980L..27I} yield a higher value at $v_{\rm sh} \sim 200 \, \text{km}\, \text{s}^{-1}$.

\subsection{Back into the void \label{sec:escape}}
The masses of the population III stars emerging from the shock-cooling pathway cannot be ascertained without being able to pin down the precise shock parameters, and run through a detailed high resolution star-formation simulation. However, it should be safe to assume that the stars are more massive than later generations (as cooling routes are still less numerous than with the higher metallically cases). Then Fig.~13 of \cite{2023ARA&A..61...65K} brackets a plausible rough temperature range of $5\times 10^4 \, \text{K} \lesssim T \lesssim 10^5 \, \text{K}$ for zero metallicity stars of masses between $20 \text{M}_{\odot}$ and $120 \text{M}_{\odot}$. 
Using Wien's displacement law $\lambda_{\rm peak} T = 2.897\times 10^{-3} \, \text{m}\, \text{K}$, the peak luminosity of the stars are thus given in the frequency range $5\times 10^{15} \, \text{Hz} \lesssim \nu_{\rm peak} \lesssim 10^{16} \, \text{Hz}$, which sits just above the ionizing frequency $\nu_{\rm ion} \approx 3\times 10^{15} \, \text{Hz}$. 
The excess energy for the peak photons is given by $h(\nu_{\rm peak}-\nu_{\rm ion})$, which converts into a proton speed (with outward directional bias following photon momentum) of $\sim 40-75 \, \text{km} \, \text{s}^{-1}$. This range is roughly comparable to the ionization shock acceleration speed computed by \cite{2004ApJ...610...14W}, and the minimal escape velocity estimate of $\sim 20 \, \text{km}\, \text{s}^{-1}$  \cite{2023ARA&A..61...65K,2004ApJ...613..631K,2004ApJ...610...14W,2006ApJ...639..621A,2007ApJ...659L..87A} needed to gravitationally bound ionized gas. 
Moreover, taking $r_{{\iota}{\rm LRD}} \sim 100\,\text{pc}$ as the representative LRD virial radius, 
an $a_0$-sized background Cosmic Tidal acceleration can additionally provide a comparable $25\,\text{km}\,{s}^{-1}$ boost over this runway, along its repelling directions, so HII gas can robustly achieve outward speeds in the tens of kilometers per second via multiple mechanisms. 

For such gas to escape the local baryonic-matter-crafted gravitational potential well, i.e., for the proton speeds to exceed the escape velocity, a very loose (assuming no help from supernovae, or accepting a smaller gas population associated with only the higher energy tail) mass ceiling on the order of $10^7-10^8\,\text{M}_{\odot}$ obtains. 
These values are smaller than the $10^{9}-10^{11}\,\text{M}_{\odot}$ stellar mass range (see e.g., \cite{2024arXiv240610341A,2024ApJ...969L..13W}) for the evolved stars or AGN interpretations of LRDs, but are not directly comparable as matter largely pass through the $\iota$LRDs but station permanently with those alternatives. 
Nevertheless, at any moment, there is only so much matter allowed within an $\iota$LRD, so there may be constrictions on how much ionization it is capable of processing. 

The number of population III stars present at any time can only be on the order of $10^{6}$, and we could pack at most a few tens to hundreds of generations of these stars into the $300$ million year interval between redshifts\footnote{Note the dark HI streams and HII outflows do not need to be confined into the reionization window, and can take longer to transport hydrogen around, so long as the hydrogen ionization states do not change while on the journeys.} $6$ and $8$. Each star puts out $\sim 10^{63}-10^{64}$ LyC photons over its lifetime as per Tb.~A1 of \cite{2023ARA&A..61...65K}, so each $\iota$LRD could potentially ionize all the neutral hydrogen within a maximum volume of $\sim 4\times 10^{-3}/n_{\rm HI}$ ($20 M_{\odot}$ stars) to $5\times 10^{-2}/n_{\rm HI}$ ($120 M_{\odot}$ stars) $\text{Mpc}^{3}$, where $n_{\rm HI}$ is the hydrogen atom number density in units of $\text{cm}^{-3}$. The intergalactic particle density is very roughly on the order of one particle per cubic meter (today's value) times $(1+z)^3$, thus $n_{\rm HI} \sim 10^{-3} \, \text{cm}^{-3}$, so even the more massive stars can only account for a fraction of the reionization workload if the number density of LRDs is around $10^{-4}$ to $10^{-3}$ $\text{Mpc}^{-3}$ (see e.g., \cite{2025ApJ...986..126K} and references therein). Unless many more (e.g., dimmer or with variant spectral shapes to selection templates) $\iota$LRDs remain to be found, which is plausible given their large distances to us, other mechanisms for reionization may be more dominant.

\section{V-shapes upright and inverted \label{sec:spec}}
The bremsstrahlung (free-free) radiation within HII regions are treated in standard textbooks such as \cite{2016era..book.....C} and we merely\footnote{Note though the energy for $\sim 0.5\, \mu\text{m}$ photons is comparable to the electron kinetic energies at just above $10^4\, \text{K}$, so HII regions need to be much hotter than this for the small deflection angle assumption to be valid. This is not the case for our context, so strictly speaking higher order corrections are needed, and we obtain only zeroth order approximates. Other caveats include that the massive particles in the HII region being assumed to be in local thermal equilibrium, which may not be guaranteed if the HII gas is engaging in rapid outflow, in which case there may not be sufficient time for thermalization.} extend the computations beyond the radio regime. The low frequency simplification for radio comes in when computing the energy output spectrum of single encounters, which can be taken as being simply flat as the radio wavelengths are far above the minimal timescale cutoff at $\lambda_{\rm min} \approx 2\pi b c /v$ ($b$ being the impact parameter, $v$ and $c$ the speeds of electron and light). For the rest-optical emission surrounding $\sim \, \mu\text{m}$ however, we land right in the region near $\lambda_{\rm min}$ (e.g., with electron temperature $T_e \sim 10^4 \,\text{K}$, we have $\lambda_{\rm min} \sim 3 \mu\text{m}$ \cite{2016era..book.....C}), so we have to properly Fourier-transform the single encounter temporal power output profile, yielding a red corrective cutoff factor (after ignoring a negligibly small term)
\bea
\zeta^P_{\nu} \approx \exp\left(-\frac{b \nu }{v}\right) \left(\frac{b  \nu}{v} + 1 \right)\,,
\eea
to be multiplied onto the constant spectrum otherwise derived in the same way as for the low frequency cases. 

The next step towards the emission coefficient is an integration against the Maxwellian $v$ profile, which yields ($G$ being the Meijer G-function)
\begin{align}
\zeta^{\epsilon}_{\nu} = 
\frac{b^2 \nu ^2 \xi}{4 \sqrt{\pi }}  \bigg[&G_{0,3}^{3,0}\left(\frac{1}{4} b^2 \nu ^2 \xi \Big|
\begin{array}{c}
 -1,-\frac{1}{2},0 \\
\end{array}
\right)
\notag \\
+2 &G_{0,3}^{3,0}\left(\frac{1}{4} b^2 \nu ^2 \xi \Big|
\begin{array}{c}
 -\frac{1}{2},0,0 \\
\end{array}
\right)\bigg]\,,
\end{align}
where $\xi \equiv m_e/(2k_B T_e)$. 
One then needs to integrate over the impact parameter $b$. We note that $\zeta^{\epsilon}_{\nu}$, when seen as a function over $b$, is also a Heaviside step function style (only smoother) cutoff for higher $b$ values, which creates a $\nu$-dependent effective upper integration limit $b_{\rm max}$. This value does not matter though, because the impact parameter's contribution to the integrand is mainly through a $1/b$ factor that drops rapidly already at large $b$ values, so cutting it off a little earlier or later does not materially impact the integration result. As such we could ignore the dependence on $\nu$ and swap in a constant $b_{\rm max}$. The integration over $b$ then proceeds as with the standard radio case, yielding a multiplicative factor of $\ln(b_{\rm max}/b_{\rm min})$. 

\begin{figure}[tb]
  \centering
\begin{overpic}[width=0.9\columnwidth]{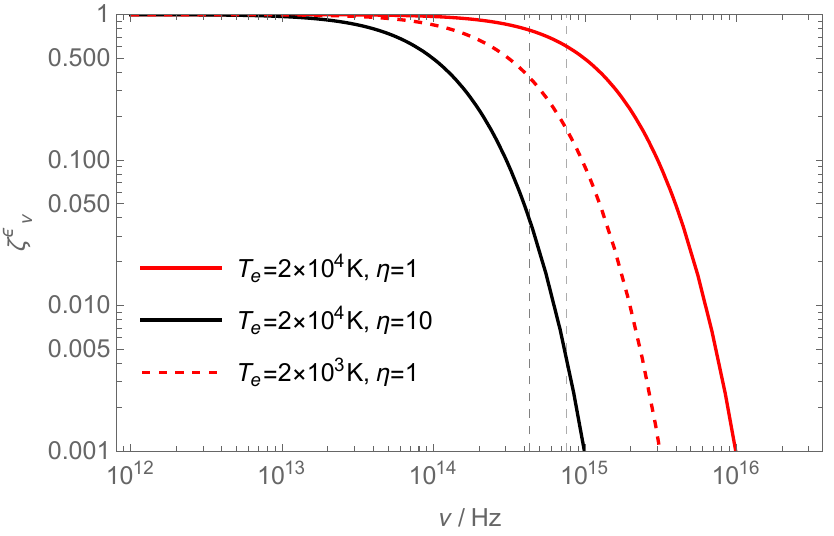}
\end{overpic}
  \caption{The corrective cutoff factor $\zeta^{\epsilon}_{\nu}$ against photon frequency $\nu$. The three curves correspond to different $T_e$ and $\eta$ parameter choices. The vertical dashed grey lines tag wavelengths $0.4\, \mu\text{m}$ and $0.7\, \mu\text{m}$, bracketing the rest-optical regime. 
}
	\label{fig:CorrectiveFactor}
\end{figure}

The $1/b$ dependence of the integrand also allows us to swap the $b$ inside $\zeta^{\epsilon}_{\nu}$ with the $b_{\rm min}$ value that dominates the integral. 
There are some textbook standard\footnote{It is also standard practice to ignore the $v$ dependence in $b_{\rm min}$, which should have complicated matter when integrating over $v$.} arguments for estimating $b_{\rm min}$, such as setting $b^c_{\rm min} = e^2/(m_e v^2)$, which corresponds to the classical maximum momentum transfer that results in the electron bouncing back in full; or letting $b^q_{\rm min} = \hbar/(m_e v)$, as required by quantum uncertainty. For the parameter range relevant for us, we have, e.g., $b^c_{\rm min} \sim 10^{-18} \, \text{cm}$ and $b^q_{\rm min} \sim 10^{-7} \, \text{cm}$ for $v_{\rm rms}^2 = 3 k_B T_e / m_e$ at $T_e \sim 10^4 \, \text{K}$. The quantum $b^q_{\rm min}$ is more stringent, so we adopt it, but introduce a fudge factor $\eta$ to account for the fact that $m_e v$ is the overall momentum rather than more appropriately its uncertainty, so in principle $b_{\rm min}$ should be larger than $b^q_{\rm min}$. In other words, we set $b_{\rm min} =  10^{-7} \eta \, \text{cm}$ with $\eta \geq 1$. Increasing $\eta$ and/or lowering $T_e$ shift the cutoff to lower frequencies, as demonstrated in Fig.~\ref{fig:CorrectiveFactor}.

For lower frequencies, the HII plasma is opaque with opacity $\tau \gg 1$, so one observes thermalized blackbody photons, initially supplied by bremsstrahlung, but brought into thermal equilibrium with the massive particles through the many collisions on their way out. This frequency range therefore exhibits a Planck's law spectrum. The temperature of the blackbody though, is of the outermost surface where $\tau$ drops below unity, which could well be lower than the inner regions where bremsstrahlung radiation originated. 
For $\tau \ll 1$ on the other hand, the HII plasma is transparent, so one is able to see the original bremsstrahlung radiation (emitted by electrons assumed to be in local thermal equilibrium, but the photons themselves are not, as they decouple from the charges after birth), displaying a flat spectrum multiplied by our reddening corrective factor $\zeta^{\epsilon}_{\nu}$. 

We can crudely estimate the temperature of the population III stars' HII regions by invoking the Wien's displacement law again, taking $\nu_{\rm peak} - \nu_{\rm HI}$ (for the $5\times 10^4\, \text{K} \lesssim T \lesssim 10^5\, \text{K}$ stars that we adopted earlier) as the peak frequency of a HII region blackbody spectrum, which gives  $2\times 10^4\, \text{K} \lesssim T_e \lesssim 7\times 10^4\, \text{K}$, overlapping the typical temperature range provided by e.g., \cite{2004ApJ...610...14W} on its lower end, and so we will take $T_e \sim 2\times 10^4 \, \text{K}$ from here on. We can then plug this $T_e$ value into the expression for opacity \cite{1967ApJ...147..471M} ($\Phi$ denoting emission measure)
\begin{align}
\tau \approx \, & 3.28\times 10^{-7} \left(\frac{T_{e}}{10^4 \, \text{K}}\right)^{-1.35} \left(\frac{\nu}{\text{GHz}}\right)^{-2.1} 
\times \notag \\
&\times
\left(\frac{\Phi}{\text{pc} \, \text{cm}^{-6}}\right)\,,
\end{align} 
which although formulated without the reddening factor, applies to the longer wavelength side of the spectrum, from which we can approach the inflection point where $\tau \approx 1$. Reading off the maximum thermal-applicable frequency location at $\sim 4\times 10^5\, \text{GHz}$ from Fig.~13 of \cite{2025arXiv250710659L} for J1025+1402 (we lean heavily on this low redshift LRD in this section, as its proximity affords much more detailed and diverse data, even though its properties might possibly be non-identical to its higher redshift counterparts), we obtain an emission measure $\Phi \sim 4\times 10^{18} \, \text{pc} \, \text{cm}^{-6}$. Once again, taking $r_{{\iota}{\rm LRD}} \sim 100\, \text{pc}$ as a typical size for $\iota$LRDs, we obtain an ``average''\footnote{Note this is not the usual volume average, or the total HII mass would already exceed the mass ceiling imposed in Sec.~\ref{sec:escape}. Its square is also much larger than the volume average of $n^2_e$ that would appear in luminosity computations, because the 1-dimensional line average puts much less weight onto the outer low density regions than a 3-dimensional volume average would. Naively plugging $\bar{n}^2_e$ into the $\lambda L_{5100}$ computation yields a much larger value than observed.} electron number density of $\bar{n}_{e} \sim 2\times 10^8\, \text{cm}^{-3}$. This $\bar{n}_{e}$ value is far above the nearby later-type stellar HII regions, but not unfathomable given that the hydrogen gas within $\iota$LRDs are supposed to exist in a pre-stellar configuration, engaging in vigorous starbursting. Noting that the monotonicity dependences of $\tau$ on $\nu$ and $\Phi$ are the opposite, it is then natural that the much larger $\bar{n}_{e}$ and $r_{{\iota}{\rm LRD}}$ pushes the $\tau =1$ threshold up into the infrared regime for $\iota$LRDs, as compared to radio for late-type stellar HII regions. 

The high $\bar{n}_{e}$ estimate is also in broad agreement with the hydrogen number density estimate obtained from [Fe II] lines, and the higher-than-AGN H$\alpha$/H$\beta$ line ratios observed \cite{2025arXiv250710659L}. As well, a local peak HI density (assuming broadly comparable to HII) at $\gtrsim 10^{9} \, \text{cm}^{-3}$ has been argued to be necessary to produce the absorption features in the Balmer lines and the strong Balmer breaks \cite{2024MNRAS.535..853J}. Because the $n=2$ state of hydrogen is short-lived thus hard to come by, collisional excitation within extreme densities that can saturate out the $n=2$ abundance fraction at the collisional equilibrium value is therefore desired \cite{2025ApJ...980L..27I}. 

We also attempt to fit the spectrum shape, and examine in particular whether $\zeta^{\epsilon}_{\nu}$ can match the redness of the observed rest-optical sector. As well, it had been shown by \cite{2025arXiv250710659L} that the rest-infrared can be fitted to a blackbody at $5000\, \text{K}$. We reproduce this fit and note also that it makes sense that this temperature is just below $T_{\alpha}$, because the outer surface of HII regions border the HI regions beyond, which have not been engaging in star formation, thus should have been cooled only by atomic cooling to just below $T_{\alpha}$, without the runaway hydrogen molecule formation setting in to lower the temperature further to hundreds of Kelvin. The transition region between the rest-optical and infrared regimes requires more detailed radiative transport treatment beyond the scope here, so we do not try to fit this part. 
For the star-generated ultraviolet sector, we also apply a blackbody fit, and since the spectrum is flatter than the temperature-insensitive Rayleigh-Jeans approximation, we expect the peak to not be very far to the left of the observation window, and so the fit should yield an estimate on the temperature of the stars. 

\begin{figure}[tb]
  \centering
\begin{overpic}[width=0.9\columnwidth]{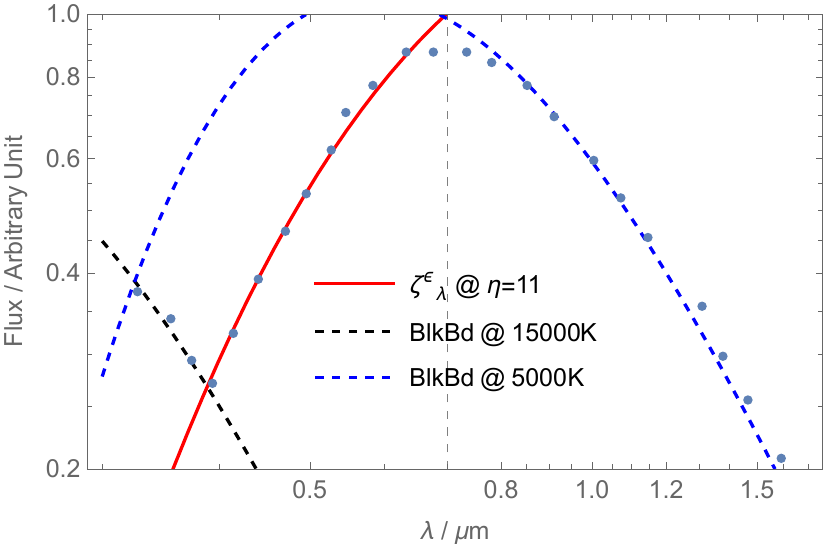}
\end{overpic}
  \caption{
The fitting to the spectral energy distribution of J1025+1402, against wavelength $\lambda$, is executed by inspection, yielding $\eta = 11$ for the bremsstrahlung radiation curve in the rest-optical sector, as well as $T=15000\,\text{K}$ and $5000\,\text{K}$ for the Planck curves in the ultraviolet and infrared sectors respectively. The overshooting inverted V-shape near the peak wavelength at $0.7\, \mu\text{m}$, marked by the vertical grey dashed line, is due to the opacity increasing so the higher temperature bremsstrahlung spectrum should transition into a lower temperature blackbody one. Our simple calculation without detailed radiation-transport is not capable of fitting to this segment of the spectrum. In the interest of demonstrating that the fit must be segmental, we do not cut off the curves beyond their applicable or commanding wavelength ranges. 
}
	\label{fig:SpectrumFit}
\end{figure}

The observational spectrum to be fitted is that of Fig.~13 of \cite{2025arXiv250710659L}, for J1025+1402, down-sampled to extract only rough continuum trends for our simple demonstration-of-principle exercise, and since it is expressed in wavelength we need to swap all $\nu$ with $c/\lambda$ and multiply the power densities (need to be per $\mathring {\mathrm A}$ rather than per Hz) with an overall $c/\lambda^2$ factor. The fitting result is presented in Fig.~\ref{fig:SpectrumFit}. The rest-optical sector is fitted with $\eta=11$, that gives $\zeta^{\epsilon}_{\nu_{\rm opt}}\approx 0.01$, deep into the cutoff regime, explaining the steep redness of the rest-optical emission. The fitting to the ultraviolet sector yields a stellar temperature of around $15000\, \text{K}$, which would correspond to small $\sim 2 M_{\odot}$ low-metallicity, or $\sim 5 M_{\odot}$ solar-metallicity, stars \cite{2023ARA&A..61...65K}. This would be reasonable if the ultraviolet emission from the massive population III stars in the center of $\iota$LRDs is largely obscured by the surrounding thick layers of dense HI gas, and other substances such as some amount of dust, and the ultraviolet light showing up in our observation window is instead from the outer regions, where smaller population II stars are able to form with the help from metals enriched by previously exploded population III stars. There is also a selection bias, namely the observation window is for the lower energy ultraviolet light, lying deep into the severely curtailed Rayleigh-Jeans tail of higher temperature more massive stars, but is closer to the Planck peak for lower temperature ones. Consequently, even without attenuation, the more massive population's contribution would still be comparatively disadvantaged. 

\section{Conclusion}
Future more sensitive observations, e.g., on Ly$\alpha$ absorption \cite{2018ApJ...864...53E} and emission \cite{2020ARA&A..58..617O}, kinetic Sunyaev-Zeldovich effect \cite{1980ARA&A..18..537S}, fast radio burst dispersion measure \cite{2021MNRAS.502.5134B}, or 21cm signals \cite{2019arXiv190912491T}, may be able to peek more clearly into the reionization era, and offer up fluctuation maps containing much greater details than a single ionization fraction number averaged across the entire Universe could. Our proposed churning of gas in-and-out of compact processing centers implies much different juxtapositions of HI and HII regions, away from the sharp-bordered Str\"omgren-type HII bubbles surrounding galaxies, floating in a rather uniform sea of HI.  
It could therefore be verified or refuted observationally, by comparing with the morphological characteristics of the various regions. 

The numerous $\iota$LRDs are eventually quenched as sites of intense massive-star formation that vigorously blow HII gas out, but we do not observe disruption events, so they must evolve into something else. Their compactness and metal pollution by population III stars make them rather attractive candidates, as the progenitors of those bulge-dominated galaxies that we observe today. To chart a way towards this possible future for extinguished $\iota$LRDs, we begin by noting that once an abundance threshold is crossed, the metals could trigger more rapid cooling, collapse, and fragmentation within the gas still being carried in by the Cosmic Tide, thereby switch on the more efficient later-type star formation pathways. The new material is then quickly locked down inside long-lived stars (with less photo-heating prowess thus less able to expel matter as well), and adds to the galactic mass budget instead of being blown back out. The $\iota$LRD remnants therefore grow rapidly during the initial postmortem spurt (cf., e.g., \cite{2025arXiv250704011B}), thanks to still being fed by the gas streams. 
This scenario initiates a core-first inside-out growth path for galaxies (consistent with observed stellar age trends \cite{Lang_2014,Suess_2020}), that is in consonance with observations indicating core stars formed at $z \gtrsim 5$ in a short burst \cite{2005ApJ...621..673T,2024ApJ...969L..13W}. 

\newpage
\acknowledgements
This work is supported by the National Natural Science Foundation of China grant 12021003, and National Key Research and Development Program of China grant 2023YFC2205801. 

%\bibliography{../../apj-jour,../../References}
\bibliography{LRDs.bbl}

\end{document}